\documentclass[twocolumn,prl,showpacs]{revtex4}%
\usepackage{graphicx}%
\usepackage{amsmath}%
\usepackage{amsfonts}%
\usepackage{amssymb}
\usepackage{bm}

\def\k{{ {\bm k} }}

\def\w{{\omega}}
\def\a{{\alpha}}
\def\b{{\beta}}

\begin{document}
\title{
Reply to Comment on "Violation of Anderson's Theorem for the sign-reversing
s-wave state of Iron-Pnictide Superconductors" [arXiv:1012.0414] by Y. Bang}
\author{Seiichiro \textsc{Onari}$^{1}$,
and Hiroshi \textsc{Kontani}$^{2}$}
\date{\today }

\begin{abstract}
We explain that the study of impurity effect in iron pnictides
 \cite{Onari} are correctly calculated based on the $T$-matrix
 approximation, contrary to the Comment by Bang\cite{bang-comment}.
The replacement $\hat{T}^{\rm b}$ with $\hat{T}^{\rm b}-\hat{I}^{\rm b}$ proposed 
by Bang breaks the perturbation theory and is therefore erroneous.
\end{abstract}

\address{
$^1$ Department of Applied Physics, Nagoya University and JST, TRIP, 
Furo-cho, Nagoya 464-8602, Japan. 
\\
$^2$ Department of Physics, Nagoya University and JST, TRIP, 
Furo-cho, Nagoya 464-8602, Japan. 
}
 

\sloppy

\maketitle
In Ref.\cite{Onari}, we studied the nonmagnetic impurity effect in the
multiorbital model for iron pnictide superconductors. 
In the sign-reversing $s$-wave state ($s_{\pm}$), we found that (i)
$T_c$ is substantially suppressed by the inter-band impurity scattering, 
since the $T$-matrix has large inter-band matrix elements. (ii) This
result holds even in the unitary limit, contrary to the fact that (iii)
inter-band scattering vanishes in the unitary limit if the bare impurity
potential in the band basis ${\hat I}^{\rm b}$ is a constant matrix and
$\det\{{\hat I}^{\rm b}\}\ne0$ \cite{Senga}.
In iron pnictides, the statement (ii) holds since 
${\hat I}^{\rm b}$ has large $\k$-dependence.

In Ref. \cite{bang-comment}, Bang complained that the result (iii) is
incorrect, that is, he claimed that inter-band scattering exists in the
unitary limit even if $\det\{{\hat I}^{\rm b}\}\ne0$. 
Here, we explain the formalism of the $T$-matrix approximation 
when ${\hat I}^{\rm b}$ is constant,
and point out the error in Ref. \cite{bang-comment}.
The $T$-matrix for a single impurity in the band-basis is 
\begin{eqnarray}
{\hat T}^{\rm b}= ({\hat 1}- {\hat I}^{\rm b} {\hat g}_{\rm loc}^{\rm b})^{-1} {\hat I}^{\rm b}, 
\label{eqn:Tb2}
\end{eqnarray}
where ${\hat g}_{\rm loc}$ is
the local Green function.
In the $T$-matrix approximation, the normal and anomalous self-energies 
for dilute impurity concentration $n_{\rm imp}\ll1$ are given as
\begin{eqnarray}
{\hat \Sigma}^{\rm n}(i\w_n)&=&n_{\rm imp}{\hat {T}}^{\rm b}(i\w_n),
 \\
{\hat \Sigma}^{\rm a}(i\w_n)&=&n_{\rm imp}{\hat {T}}^{\rm b}(i\w_n)
\hat{f}(i\w_n){\hat {T}}^{\rm b}(-i\w_n),
 \label{eqn:Self}
\end{eqnarray}
where $\hat{f}(i\w_n)\equiv\sum_{\k}\hat{F}_{\k}(i\w_n)\ll 1$
is the linearized local anomalous Green function.
${\hat \Sigma}^{\rm a}$ is diagrammatically expressed in Fig. 1 (a),
which expresses the inter-band scattering of a Cooper pair
when ${\hat {T}}^{\rm b}$ has off-diagonal elements,
as shown in Fig. 1 (b).
In the $s_\pm$-wave state, this inter-band impurity scattering
suppresses the superconductivity.

\begin{figure}[!htb]
\includegraphics[width=0.9\linewidth]{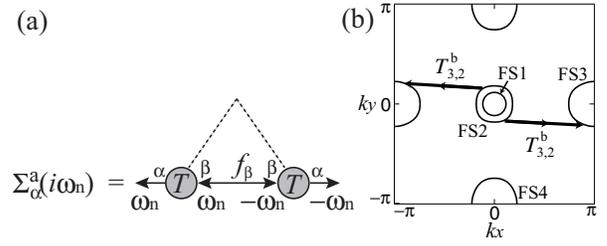}
\vspace{-3mm}
\caption{(a)Diagrammatic expression for ${\hat \Sigma}^{\rm a}$.
(b)Inter-band scattering of a Cooper pair 
described by ${\hat \Sigma}^{\rm a}$.
}
\label{dia}
\vspace{-5mm}
\end{figure}
When ${\rm det}\{{\hat I}^{\rm b}\}\ne0$,
eq. (\ref{eqn:Tb2}) becomes
$T_{\a,\b}^{\rm b} = -1/g_{{\rm loc},\a}^{\rm b}\cdot \delta_{\a,\b}$
in the unitary limit ($I\rightarrow \infty$),
which is band-diagonal even if ${\hat I}^{\rm b}$ is not band-diagonal.
Therefore, the pair-breaking due to inter-band scattering 
is absent in the unitary limit unless ${\rm det}\{{\hat I}^{\rm b}\}=0$
  \cite{Senga}.

On the other hand, Fe-ion substitution in iron pnictides
induces the orbital-diagonal local impurity potential.
Then, ${\hat I}^{\rm b}$ is given as
${\hat I}^{\rm b}_{\k,\k'}= I {\hat U}^\dagger_\k {\hat U}_{\k'}$,
where ${\hat U}_{\k}$ is the transformation matrix
between orbital- and band-bases.
Because of the large $\k$-dependence in iron pnictides,
${\hat T}^{\rm b}$ is not diagonal even in the unitary limit,
and therefore $s_\pm$-wave state is fragile against impurities.
This is the main result in Ref. \cite{Onari}.

In Comment\cite{bang-comment}, Bang seems to claim that the bare
impurity potential $\hat{I}^{\rm b}$ should be subtracted from the
$T$-matrix $\hat{T}^{\rm b}$. If we follow his comment in eq. (3), 
inter-band scattering of Cooper pair always occurs.
However, such subtraction violates the perturbation
theory, and induces various unphysical results, for example,
the divergence of $\hat{\Sigma}^{\rm a}$ for
$I^{\rm b}\rightarrow\infty$ will give the divergence of $T_c$ 
in the $s_{++}$ wave state. We agree that the real-part of
the ``normal self-energy", which becomes $n_{\rm imp} \hat{I}^{\rm b}$
in the Born limit \cite{comment}, is absorbed by the change in the chemical
potential. However, this fact never means that $T$-matrix is renormalized
to $\hat{T}^{\rm b}-\hat{I}^{\rm b}$, contrary to the claim by Bang.

In the constant $\hat{I}^{\rm b}$ model, Bang\cite{bang-comment} showed
that the $T$-matrix is not band-diagonal when $\hat{I}^{\rm
b}\propto\left(\begin{array}{cc}1&1\\ 1&1\end{array}\right)$, ({\it i.e.} $\det\{\hat{I}^{\rm b}\}=0$) in the unitary limit. However, this is the special case (measure
zero probability) in the statement (iii). In contrast, $\hat{T}^{\rm b}$
always has off-diagonal elements in multiorbital models since
$\hat{I}^{\rm b}$ is not constant. Thus, it is better to analyze the multiorbital model for a quantitative study of impurity effects in iron pnictides.

In summary, our studies of impurity effect in iron pnictides\cite{Onari}
are correctly calculated based on the $T$-matrix approximation, which
becomes exact in the dilute limit. The replacement $\hat{T}^{\rm b}$ with
$\hat{T}^{\rm b}-\hat{I}^{\rm b}$ in eq. (3), which was proposed by Bang
\cite{bang-comment}, breaks the perturbation theory and is therefore erroneous.



\end{document}